# Unpacking Adherence and Engagement in Pervasive Health Games

**Magy Seif El-Nasr[1], Shree Durga[2], Mariya Shiyko[3], and Carmen Sceppa[3]**
[1]Colleges of Computer and Information Science and College of Arts, Media and Design
[2]Plait Lab, College of Arts, Media and Design
[3]Bouve Health Sciences College
Northeastern University, 360 Huntington Avenue, Boston, MA 02115
{magy, sh.subramanian, m.shiyko, c.sceppa}@neu.edu

## ABSTRACT
Pervasive health games have a potential to impact health-related behaviors. And, similar to other types of interventions, engagement and adherence in health games is the keystone for examining their short- and long-term effects. Many health-based applications have turned to gamification principles specifically to enhance their engagement. However, according to many reports, only 41% of participants are retained in single player games and 29% in social games after 90 days. These statistics raise multiple questions about factors influencing adherence and engagement. This paper presents an in-depth mixed-methods investigation of game design factors affecting engagement with and adherence to a pervasive commercial health game, called *SpaPlay*. We analyzed interview and game behavior log data using theoretical constructs of sustained engagement to identify design elements affecting engagement and adherence. Our findings indicate that design elements associated with *autonomy* and *relatedness* from the Self-Determination Theory and *integrability*, a measure of how well activities align with a person's life style, are important factors affecting engagement and adherence.

## ACM Classification Keywords
H5.m. Information interfaces and presentation (e.g., HCI): Miscellaneous.

## INTRODUCTION
User engagement is an important construct that many designers aspire for their products. There has been extensive work on modeling, defining, and understanding engagement from multiple perspectives and fields. For example, the field of business has been concerned with understanding engagement to measure and tune marketing and business goals [5, 6]. Education researchers focused on engagement within the learning context [57]. Human Computer Interaction (HCI) researchers have also been interested in defining engagement. For instance, O'Brien and Toms [39] defined engagement in terms of challenge, positive affect, endurability, aesthetic and sensory appeal, attention, feedback, variety/novelty, interactivity, and perceived user control. Other work identified factors, such as attention and curiosity [2], flow [15] and intrinsic motivation [6, 7].

While much research has been devoted to defining engagement, little work involved empirical studies to understand the design properties that affect engagement. Recently, there has been some excitement that gamification, using gaming elements in non-gaming products [5], can lead to increased engagement [13]. In addition, there has been several review studies evaluating the use of games for health specifically for rehabilitation and older adults [5, 27, 28, 34, 36]. However, despite the many studies, and also the ones specifically targeting game engagement [3, 5, 7, 9, 24, 33], we still face an open problem of understanding game design elements that affect engagement. Furthermore, for health applications engagement alone is not sufficient. Health behavior change interventions, in particular, require consistency and participants sticking to a routine for a period of time – i.e., longitudinal adherence. In the medical and health fields, the term adherence denotes how much users adhere to a medication or in this case the game prescribed for health benefits. Thus, the overarching question this paper seeks to address is what factors in a game's design affect engagement and adherence? We define engagement behaviorally through the activities and the time invested in the game, and adherence as the consistency of logging in to play over time. This study has implications on health and educational applications [4].

This is a direly needed work, because data show that games vary in terms of their engagement and adherence levels. According to Flurry, a mobile analytics company, for iOS and Android apps only 54% of the customers return after 1 month, 42% after 2 months, and 35% after 3 months [43]. Earlier report [42] shows that in 2011 user retention was only 14% for 6 months and 4% for 12 months. Game applications do not hold highest user retention value, compared to other applications, specifically, single player games tend to retain 41%, while social games retain 29% after 90 days [42].

In this paper we investigate the level of adherence and engagement within *SpaPlay*, a game designed to encourage healthy eating and exercise. *SpaPlay* is a pervasive health game similar to [22] developed to induce behavior change. We chose this game due to access of the game's data and the cooperation of the game company to facilitate use of data and adjustment of the design based on our study. In this study, we use a mixed-methods approach to examine the effect of playing the game on health behavior change. To that end, our study incorporates both qualitative analysis of participant interviews and quantitative analysis of logged gameplay data to examine factors affecting adherence and engagement with the game. In particular, we aim to look at factors discussed by previous work, such as motivational factors using Self Determination Theory, the game's integrability to life style, game mechanics, and gameplay experience [26, 28, 29, 6, 22, 42, 45]. These constructs were used to code the interview data allowing us to quantify how much of the participants' experience with the game is related to each one of these factors. Gameplay activity logs allowed us to measure engagement (i.e., how much time participants spent in the game and how much activities they did over multiple sessions) and adherence (i.e., how much consistency in logging into the game did participants exhibit within a period of time).

The paper is divided into the following sections. First, we review the theoretical work that our analysis utilizes, including Self Determination Theory (SDT) of intrinsic motivation, Integrability theory, and Game Design theory. Second, we review some of the most relevant previous work. Third, we discuss the game. We then discuss the study and results. We conclude with a discussion section unpacking game design factors affecting engagement and adherence constructs within the context of playing *SpaPlay*.

## THEORETICAL CONSTRUCTS FOR SUSTAINED ENGAGEMENT

When it comes to engagement and adherence in games, there are a handful of frequently referenced successful examples, such as *Lineage* or *World of Warcraft*. In fact, many of player engagement and motivation theories have been drawn from observations of a chosen handful of games [e.g., 37]. A popular theory that has been used to explain and examine factors for sustained engagement is SDT of intrinsic motivation. SDT frames engagement in terms of basic psychological needs, such as needs for *autonomy, competence* and *relatedness* [17, 18]. In Self Determination Theory (SDT), *competence* is defined as the innate desire to grow our abilities and gain mastery of new situations and challenges; *autonomy* as the need to take action out of personal volition and not because we are controlled by circumstances or by others; and *relatedness* as our need to have meaningful connections to others. Competence as a construct also overlaps with *flow*, a cognitive state of full involvement and immersion which usually is observed when skill matches difficulty of the task [16]. As discussed above, we focused on the constructs derived from SDT rather than *flow* as a theory for our study. It should be noted that designs that apply high autonomy, relatedness and competence (constructs of SDT) have shown to have a great effect on behavior change [22, 41, 44].

In addition, the concept of *integrability to lifestyle* is a critical construct investigated within the domains of consumer behavior and market research, particularly exploring connection of lifestyle measures, such as activities and behavior, values and attitudes, to how it influences individuals' use of leisure and technology [26, 28, 29]. It is also an important construct that has informed research in health behavior change, specifically in designing interventions to facilitate management of chronic health conditions, such as diabetes, or weight-loss [10, 14, 15]. Health intervention studies have found that patients were more likely to adhere to health interventions when the goals aligned to "lifestyle goals", such as goals tailored to individual participant's physical activity and weight-management goals, sensitive to cultural and ethnic values and norms [26]. More recent studies on pervasive health applications for behavior change have found that incorporating lifestyle-based intervention strategies would greatly benefit users who face motivational challenges [4, 23, 33]. However, within serious games, there is paucity in research exploring the concept of lifestyle with engagement and adherence. This paper is an early attempt to use empirical player data to establish connections between lifestyle integrability, and engagement and adherence in pervasive health games.

Lastly, the Mechanics, Dynamics, Aesthetics (MDA) model developed by Hunicke et al. [32] discusses a theoretical framework for understanding the process of game design. Hunicke et al. defined game mechanics as the rules of the game system, dynamics as the systems that emerge from these rules interacting at playtime, and aesthetics as the users' experience of the dynamics of the game. In this paper, we have incorporated game mechanics and game play experience (aesthetics) as constructs to code our interview data. Both these factors have been discussed to have significant effect on engagement. For example, studies have shown feedback, narrative, pacing to have an effect on engagement, immersion, and presence [7].

We acknowledge that the discussion here is not complete, since there are many other theories discussed in both game design and HCI about factors affecting engagement and adherence. However, in this paper, we decided to use constructs from these four theories as a start since they have shown great importance based on previous work. Subsequent empirical studies are needed to expand beyond these theories.

## PREVIOUS WORK

Beyond the theoretical conceptualization, there has been very few works measuring engagement, and even less empirical studies examining engagement and adherence. Mekler et al. [35] reviewed 87 quantitative studies on game enjoyment; they found only 3 studies examined the phenomenon outside of the lab and none of the studies were longitudinal, i.e. addressing the topic of adherence over days or months of play. For pervasive games, lab studies are of limited utility to understand how players engage with and adhere to the game over time.

Of the few studies that have investigated sustained engagement over time is the work of Rigby et al. [47]. Rigby et al. use SDT to explain why some people adhered to certain games, such as World of Warcraft for several months. However, their study was methodologically limited to just player interviews, and thus no behavioral data was used. Building on this research, we focus on understanding how players engage and adhere to a pervasive health game experience using behavioral and interview data.

In addition, there has also been some work, in the context of health behavior change that looked at longitudinal measures of persuasion and health behavior change. For example, Munson and Consolvo [37] investigated goal setting, rewards, self-monitoring on the user experience. Results show that goal setting and self-monitoring have clear benefits. Additionally, Consolvo et al. [14] further investigated the role of goal setting and tracking on sustained engagement. While these studies provide valuable insights on design, they (a) do not target game design, and (b) are methodologically different from the work presented here. In particular, we investigate behavioral measures of engagement and adherence based on qualitative participant input on motivational, situational, and cognitive elements.

## INVESTIGATING ADHERENCE AND ENGAGEMENT

We developed a 45-day field study to investigate game design factors affecting adherence and engagement of a commercial health based game called *SpaPlay*. Game log data and interviews were used to investigate two variables – engagement and adherence. Again, we describe engagement through time participants spent in the game and activities they did over multiple sessions; adherence we define as consistency in logging into the game within a period of time.

### Pervasive Game Application – *SpaPlay*

*SpaPlay* is a pervasive health game that seeks to foster healthy eating and exercise through social network and gameplay. Figure 1 shows a screenshot of the game. The game is currently deployed on the web and will be distributed and deployed through mobile devices. Players in *SpaPlay* run a virtual spa resort whose rating depends on a combination of player activities within the game (e.g., tending to the island, watering vegetation, cleaning up virtual running-tracks) and in real-life (e.g., performing real-life physical activities, eating healthier foods).

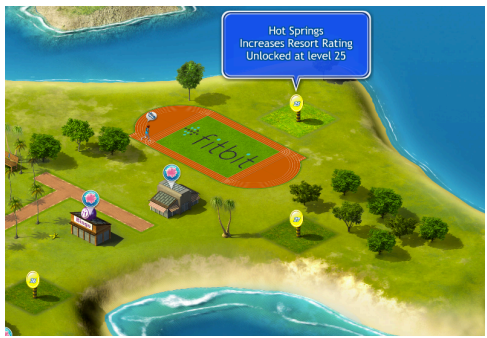

**Figure 1. Screenshot of the game**

The game has two core mechanics through which players earn experience points: *Quests* and *Sparks*. Sparks are short, single-time tasks that might include stretching for five minutes or adding spinach to a sandwich. Quests are a series or sets of physical activities or dietary tasks that the players complete within the span of a week. An example of an exercise quest is: taking one flight of stairs twice a week. Similarly, an example for a food quest is: consciously substituting a sugary beverage for water five times or eating a fruit instead of a snack twice in a week. Below we discuss the game mechanics in detail.

**The Virtual Island.** Players in *SpaPlay* run a virtual spa resort, whose rating depends on a combination of player actions or activities within the game, e.g., tending to the island, watering vegetation, cleaning up running tracks in the island, and doing real-life physical activities.

**Game Leveling up and Rewards.** Experience points in the game are accrued through activities done outside of the game, such as physical exercise or making healthy eating choices. These activities are tracked through the game using physical activity sensors or self-report (as checked sparks or quests). Once tracked, they translate to experience points that level up players' characters. Both sparks and quests are used to encourage players to develop fondness towards the game and add playfulness to ordinary day-to-day health-related activities, such as walking to the next bus stop. An example quest is 'Tame the sugar monster', which takes time and sustained persistence to complete, and thus are worth more experience points.

**Game Feedback: Extrinsic Rewarding Mechanisms and Intrinsically appealing Performance Visualization.** Leveling up, just as in other games, unlocks certain in-game assets that increase the aesthetic quality of the resort and its rating. The game also includes a similar concept to leaderboards through visualizations of player's performance against friends over time. These visualizations give feedback to the player on their performance given certain outcomes, and are designed as extrinsic rewards for activities. Each time a player logs in, there is a visualization reflecting player's activities and game-related progress, including the number of sparks and quests completed at different time intervals (e.g., past week, each day, etc.).

### The Study

We recruited 22 participants to play the game. There were no criteria for selection except that participants should be familiar and like to play social games. 4 participants dropped out of the study after the initial informational survey session due to lack of time to commit to the study. Thus, 18 participants (14 male and 4 female; 15 White and 3 Asian) comprised the final sample. Age of participants ranged from 18-22. Participants were asked to come to the lab to sign the consent form. During that time they were also given a demo of the game and asked to play it for 5 weeks. We also scheduled 20-30 minute, one-on-one interviews with participants once a week during the 6-week period in which they played the game.

### Data Collection Methods

We collected two types of data during this study: (a) game telemetry and (b) interview data. Game telemetry is a log of all game activities, such as login, logout, sparks done, quests planned, quests and sparks done with corresponding timestamps.

In addition to this quantitative log of participants' interactions with the game, we also conducted weekly 20-30 minute interviews to assess the game experience. In particular, the interview data was collected with the aim of understanding players' behaviors and their perception of their engagement with the game. We chose not to bias the data collection with any theory to allow us to look at other factors that may emerge that may not have already been identified by the theory. We used Seidman's [49] in-depth three-step interview structure with the goal of "understanding the lived experience of other people and the meaning they make of that lived experience [49]."

The three-step process is as follows:

- **Establishing life-context.** In the first interview we included questions to establish participants' *life context*, reasons they chose to play the game, and their initial thoughts about their gameplay. Example questions were: "what were your initial impressions of the game?", "what are your favorite quests and sparks in the game?", and "how frequently do you typically workout during a week?"

- **Establishing details of the experience as players see it.** The focus of the second interview was on *reconstruction* of details of participants' experiences and contexts behind these experiences. We used participants' automatically recorded game data to construct and individualize questions. Example questions are: "seems like you were able to explore sparks and quests more this week, something you said you weren't able to do much last week, like, I noticed you have done quite a few of the diet quests. What made you explore the sparks you hadn't done before?"

- **Reflection on meaning.** Seidman points out that the third and final interview is intended to put the participants in an explicitly reflective role. So in the final interview we included questions such as: "in earlier weeks you had said that pretty much logged in once every day or every other day, and so we kind of looked in to your gameplay profile a bit and it looked kind of like *eating smaller portions*, *eating breakfast*, and *drinking water* were some of the sparks that you did the most on a consistent basis. Are these things you were particularly working on? Say more about why you chose them."

All interview data were collected in audio files, which were later transcribed in text and coded for analysis as will be discussed next. We used Dedoose (a qualitative analysis software) to code interview excerpts. We coded all interviews based on theories we deemed to be prominent factors affecting sustained engagement for health behavioral outcomes. The coding process is described in detail in the following section.

## Data Analysis Methods

**Qualitative Data Processing:** We used a content analysis method [31] using pre-existing theoretical constructs. Codes are shown in Table 1. We developed codes using three key theoretical constructs (these constructs were discussed above): (a) self-determination theory of intrinsic motivation [23, 14], (b) gameplay and mechanics [1, 10], and (c) integrability to lifestyle in pervasive technologies [5, 8]. As shown in Table 1, for SDT, players' utterances were coded with labels: competence, autonomy, and relatedness, with negative/positive labels denoting when participants are discussing game features or activities that has a negative or positive value. In addition to SDT, we also coded interview data to include codes that are related to mechanics and gameplay and how they are affecting the participants' experience, again with positive and negative labels. Mechanics codes were codes that pertained to specific rules of the game, while gameplay codes were about how the rules were experienced from a gameplay experience perspective. Finally, we also coded excerpts based on comments that discuss how well the game integrates to, or not, the users' lifestyle. Two coders were asked to code existing excerpts to establish reliability of the codes. The inter-rater agreement was 92.6%.

Prior to coding, we read through the interview transcripts several times and broke down participant utterances (sentences or groups of sentences) that conveyed salient meanings, in to excerpts. Excerpts can be thought of as a data chunk in the entire transcript that captures the essence of the code (Saldana, 2009). Through the process of coding, we coded a total of 525 excerpts in 54 documents of 18 player interviews. We then counted the number of utterances for each construct, depending on if they were negative or positive comments. We then normalized all codes per participant to allow us to compare the codes among groups of participants.

**Table 1. Codes for Qualitative Interview Data**

| |
|---|
| **Positive Autonomy** |
| ▪ Allows planning goals and to be used as checklist. |
| ▪ Provides freedom to make decisions during the day. |
| ▪ Quests helped in planning ahead. |
| **Negative Autonomy** |
| ▪ Lacks ability to personalize/individualize. |
| ▪ Lacks ability to provide freedom to choose rewards. |
| ▪ Provides less control for monitoring progress. |
| **Positive Competence** |
| ▪ Encourages user to "keep oneself honest". |
| ▪ Facilitates tracking healthy activities. |
| ▪ Likes "leveling up" & rewards for repetitive activities. |
| ▪ Allows user to return to the game. |
| **Negative Competence** |
| ▪ No incentive to "push oneself beyond". |
| ▪ Lack of feedback on activities or progress. |
| ▪ Confusions with tracking mechanisms. |
| **Positive Relatedness** |
| ▪ Familiar genre or aesthetics. |
| ▪ Synergy with game achievements. |
| **Negative Relatedness** |
| ▪ Lacks sociability. |
| ▪ Lacks of personalization |
| ▪ Not the preferred genre or type of game |
| **Positive Game Mechanics** |
| ▪ Likes the quests |
| ▪ Likes repetitive activities to level up. |
| ▪ Likes the Mini-Games. |
| ▪ Likes the Sparks. |
| **Negative Game Mechanics** |
| ▪ Lacks ability to personalize quests and rewards. |
| ▪ Confused by the rules for the rewards. |
| ▪ Lacks thinking or problem solving. |
| ▪ Lacks self-expression capability. |
| **Positive GamePlay Experience** |
| ▪ Likes to play during "Down time". |
| ▪ Likes routine. |
| **Negative GamePlay Experience** |
| ▪ Needs more socializing experiences |
| ▪ Needs to *personalize* the island |
| **Positive Integrability** |
| ▪ Adopts incremental changes into current routine. |
| ▪ Variety in activities allows for choices that are close to users' life style. |
| ▪ Ability to keep mental notes of to-dos. |
| **Negative Integrability** |
| ▪ Inability to *auto-track*. |
| ▪ Needs goals/Incentives to promote consistency. |
| ▪ Inability to break existing routine. |

**Quantitative Data Analysis:** Using real-time telemetry data, adherence was quantified in three ways. First, we computed the number of days (out of total 42 possible) participants engaged with the game. Any real-time records of log-ins on a given day were used as indicators of engagement. Second, to assess regularity of adherence to the game, we computed time intervals between successive days of play for each participant. Play regularity was summarized as individual means and standard deviations (second and third indices of adherence, consecutively). For example, an individual mean of 2 days would represent an average of 2 days between plays, and an individual standard deviation of 0 would indicate consistent and non-variable interaction with the game every 2 days. In comparison, the same mean but a standard deviation of 2 would represent inconsistency in interaction with the game, with an average time between logins of 2 days albeit variable.

The last (fourth) index of game engagement was computed as the average number of in-game activities on any given day of play for every participant. Activities included the total number of time-stamped interactions and involved completion of mini games, sparks, and quests.

To examine the effect of time on adherence, we examined whether participants were less or more likely to engage with the game over the course of 5 weeks. We applied generalized mixed-effects modeling [23, 40] with a logit link function. This model is an extension of general linear models and is used with observations recorded repeatedly over time (e.g., adherence recorded every day for 5 weeks). We coded adherence as an outcome of a yes or no. Time was included as a predictor (days 1 through 41) for the model.

**Integrating quantitative and qualitative data:** To examine the bivariate relationships between four indicators of adherence and engagement described above and the 12 theoretical factors derived from qualitative data, Spearman rho rank-order correlations [31, 34] were computed. We used non-parametric alternative to Pearson product-moment correlation due to outliers and non-normality in the data. Given the small sample size, this analysis should be interpreted as exploratory, and p-values should not be interpreted literally. Instead, the magnitude of the relationship is of more direct value.

## RESULTS

### Effect of Time on Adherence

Figure 2 presents a graphical summary of the rolling number of participants engaged with the game on every study day.

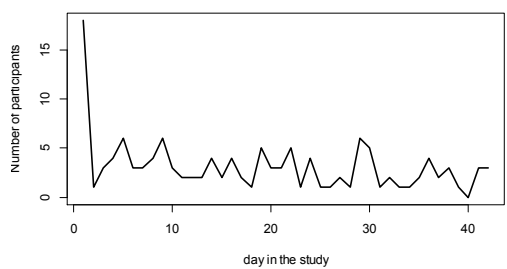

**Figure 2. Rolling frequency of participants engaged with the game across the 42 days of the study.**

Results of the generalized mixed effects modeling supported our hypotheses that adherence would reduce over time. The linear effect of time was estimated to equal .98 (on the odds ratio scale, p = .03). This implies that every day there was a 2% reduction in adherence, compared to that on the previous day. This is an average estimation of the downward trend in adherence, and the Figure 2 shows that there were some unpredictable fluctuations from day to day.

### Descriptive Statistics

Table 2 shows mean and standard deviation of the number of utterances for each category coded. The mean positive and negative utterances vary between constructs, with gameplay receiving a lot higher positive utterances than negative ones, while autonomy and integrability received similar numbers for positive and negative, and mechanics, relatedness, and competence received a lot more negative than positive utterances. The number of utterances by itself does not mean much regarding engagement or adherence, however. The positive utterances may mean that the participants engaged with that element of the game, but more negative utterances may also mean that participants engaged enough to care to discuss issues regarding this construct. Thus, deeper analysis is required to unpack engagement and adherence.

**Table 2. Spearman rho Bivariate Correlations (* p < .1; ** p < .05; *** p < .01) and Summaries of descriptive statistics. Showing mean and standard deviation of number of utterances denoted to each category.**

|  | Total Days Active | Mean Length Bt. Logins | SD of Length Bt. Logins | Level of Engagement | Mean | St. Dev |
|---|---|---|---|---|---|---|
| Positive Autonomy | .604*** | -.609** | -.638** | -.715*** | 16.67 | 12.21 |
| Negative Autonomy | -0.082 | 0.219 | 0.121 | -.580** | 19.92 | 13.11 |
| Positive Competence | 0.126 | 0.001 | -0.094 | -0.264 | 14.76 | 10.76 |
| Negative Competence | 0.027 | -0.105 | -0.203 | 0.128 | 24.83 | 24.89 |
| Positive Relatedness | 0.331 | -0.061 | 0.249 | -0.318 | 4.60 | 5.61 |
| Negative Relatedness | -0.457* | 0.464* | .599** | 0.254 | 28.77 | 25.39 |
| Positive Mechanics | 0.444* | -0.287 | -0.384 | -0.193 | 6.76 | 7.99 |
| Negative Mechanics | -0.317 | 0.134 | 0.066 | .566** | 23.31 | 17.54 |
| Positive Gameplay | 0.327 | -0.349 | -0.033 | -0.301 | 22.80 | 18.26 |
| Negative Gameplay | 0.173 | 0.007 | 0.081 | -0.146 | 7.20 | 14.15 |
| Positive Integrability | .552** | -.535** | -0.511* | -0.295 | 18.33 | 11.78 |
| Negative Integrability | -0.2 | 0.299 | 0.28 | -0.118 | 19.33 | 12.79 |

For adherence and engagement measures, we found that on average, participants engaged with the game for 7.22 days (SD = 7.05), Md = 4.5, Range = 1 to 24. This means that participants averaged 7 days of engagement out of the total of 45 days, which while comparable to other social games show very low adherence. However, the high standard deviation means that also participants varied a lot from 1 day to 24 days of engagement with the game. This alludes to the fact that for some participants some aspects of the game worked. Interviews and more analysis will help us unpack what worked and what didn't and why.

On those days when they interacted with game, participants completed an average of 39.21 in-game activities (SD = 57.12), Md = 24.2, Range = 4 to 252. This is a high number of game activities with also a large variation in the sample. This also shows the variation in play styles between participants, which deserve further analysis, as we shall see later.

We were able to compute regularity of engagement for a subsample of participants: specifically, means for 16 and standard deviations for 14 – an artifact of the days of plays available for computing necessary statistics. On average, gaps between play days equaled 7.68 days (SD = 5.23), Md = 6, Range: 1.79 to 20. Typical variability in regularity of engagement averaged at 4.29 (SD = 2.42), Md = 3.87, Range: 1.1 to 9.3.

These measures by themselves do not tell us much. However, leveraging the coded interviews and correlating these measures with the number of positive and negative utterances may allow us to make some conclusions about what elements of the game are associated with these measures, and thus adherence and engagement.

**Factors Affecting Adherence & Engagement**
Table 2 summarizes results of the bivariate correlation analyses. Significant findings are marked with asterisks. Of more importance, however, is the magnitude of the correlation coefficients. In this section we will discuss these results and outline participant quotes to contextualize the results with qualitative data.

Out of 12 factors, positive autonomy was strongly positively related to total days of play (implying that those scoring high on the number of positive autonomy utterances also had a high indicator for total days played) and negatively to mean length and SD between logins, and engagement (all $p < .05$). The relationship in this case is reversed, where higher scores on the number of positive autonomy utterances correspond to lower values on the other three indices of adherence. This means that the higher positive autonomy utterances a participant made the more engaged and the more adherence she showed to the game.

To exemplify this, we selected the following excerpt. One participant noted: "*I try not to get points for something that I did that I would've done anyway versus what I did, because I'm going to make a conscious good decision today, and do this. Like, not skipping breakfast and drinking water instead of soda are things I normally do. But on the other hand, I did add one thing to my work out. I started using a dumbbell now, and now I'm curling now just to use weights. But they are like really small weights. Nothing serious. But that is just a new thing to my routine and I try and do this everyday as my sparks*." In this excerpt the participant explains how she rewards herself for certain tasks and not for the ones she usually does. This is an important design feature in the game, because the game has numerous sparks and quests that a player can accrue points for, this design is developed to empower autonomy for players to make their own choice of how they want to reward themselves.

Negative autonomy was negatively correlated with level of engagement ($p < .05$). This means that participants who had more negative autonomy utterances tend to have less engagement with the game. As one participant states, "*But I find there is nothing keeping me in the game. One of the biggest things I feel like is missing is what I want as a reward. It seems like it would lend itself very easily to making your own Spa Resort and sort of designing where things are going to be. But instead the game is a very rigid, like do this to get to level 3 and then you get the Yoga Studio. Well, what if I don't want the Yoga Studio?*"

Negative relatedness is negatively related to total days of play and positively with mean and SD of days between logins (p values $< .1$ or $< .05$). This means that participants with higher number of negative relatedness utterances engaged less with the game and their adherence was low.

Positive mechanics was modestly related only to days of play ($p < .1$), while negative mechanics with the level of engagement ($p < .05$). This means that participants who had positive utterances towards mechanics played more while the ones who had negative utterances did less activities within the game.

Finally, positive integrability was moderately positively related to total days of play ($p < .05$) and negatively to mean and SD of days between play ($p < .05$ and .1 respectively). This, like autonomy, means that the higher positive integrability utterances a participant made the more engaged and the more adherence she showed to the game. Again this is exemplified by participants' statement, "*I am not the kind of person, who will go for a walk around the block, just because. But I will always choose to walk if I have a destination, even if it I am just going to like look at something at the store and not necessarily buy it. So for me, having a game quest wouldn't necessarily make a difference or make it interesting but it helps that now I do mostly the walking quests if I'm like going somewhere and doing something.*" This participant chose to walk to his destinations rather than take a bus or train. The option of using this in the game closely integrated to his lifestyle and thus was easier for her to stick to this routine.

## DISCUSSION
Our results show great association of Autonomy, Relatedness, Mechanics and Integrability on adherence and engagement. Autonomy had a direct effect on the number of days of play, less gaps between play, and more consistency of these gaps, reflecting higher adherence. However, positive comments about autonomy had a negative effect on time and activities done in the game (negative effect on engagement). This may be due to the fact that the game itself requires participants to do small amount of things virtually but rewards more the activities done in the real-world. Most positive utterances about autonomy in this game were related to allowing participants to choose their own routes to achieve their goals through choice of quests or sparks that are more appropriate to their own tastes.

Negative comments related to Relatedness seem to be correlated negatively with the number of days of play and positively to the gap between play sessions and its consistency, reflecting a negative effect on adherence. It may be that participants who didn't adhere to the game needed relatedness or are more motivated by relatedness than other constructs. During this study, the game did not have many quests that could be completed with friends type quests or activities. While there was an 'add your friend' feature and you can compare your scores with friends and see their activities, there were limited features that promoted the social aspect in the game. This was a factor that many participants discussed.

Integrability is another important construct that is mostly absent from studies about game engagement, but is of particular importance to pervasive behavior change games. As our results show, this construct has a direct effect on adherence. In particular, our results show that the effect of interability is significant, in that positive comments seem to be coupled with more days of play, and less days between sessions and also less variability (SD is negatively correlated) over time, and thus more adherence. We specifically argue that integrability is important for adherence. In this game, integrability was allowed through specific quests that were designed to be simple changes to someone's life style. For example, 'skip the next T-stop and walk instead', 'take the stairs to office' were sparks added to specifically allow for better intergration to someone's life style. This may be a good design construct to use – one that may require personalizing to close the gap between the person's life style and the product.

In terms of game mechanics, it is apparent that there is a direct effect of the negative comments on mechanics and the time and activities people do in the game. The main issues here were issues of too much rewards – making the game unbalanced. Thus, mechanics need to be balanced with well-structured reward system to have better audience engagement.

These results show importance of disconnecting the concepts of adherence and engagement. Interestingly, previous work alluded to many factors from SDT, integrability, flow, mechanics, gameplay, etc. as affecting engagement. Our research shows that mechanics designed to enable autonomy and relatedness (constructs from the SDT model) have a significant effect on adherence but not much on engagement. However, that may be due to the actual design of the game that requires users to do things in real-life more than virtually. Our results also show the impact of integrability on adherence. This is an important construct for pervasive applications and in particular health based pervasive games. Our results also show the importance of mechanics on engagement.

## LIMITATIONS AND FUTURE WORK

It should be noted that the current study has several limitations. First, the study is limited to 18 participants. This is a small number to draw general conclusions from. Thus, the results discussed need more studies to verify them with bigger groups. We chose to start with a small number to allow us to adjust and test the methodology with a field study. Future work will include a study involving 67 participants with a longer duration of play. Also, the study is constrained by the use of only one game for health as a case study. This is due to access and our ability to get a health game with instrumented data for the analysis needed. While similar issues of adherence have been discussed of other games, further work is needed to explore data collection from different health apps and games to generalize the results discussed here.

## CONCLUSION

Applications of pervasive games seek to transcend market numbers (reviews or sales, e.g.), and support and extend goals of the institutions in various subject areas, like healthcare, politics, education or advertising [6]. A significant amount of work has been done in the field of player engagement, examining factors such as intrinsic motivation, flow and attention [10, 25, 32]. However, there has been little work on empirically assessing the effect of such factors within a pervasive game where engagement and adherence are required. This study aims to start investigating these factors within a specific game for health. While constrained by investigating only one game, it nevertheless showed that autonomy, relatedness, and integrability are important factors affecting adherence, and game mechanics as a factor affecting initial engagement with the game. More work is needed to investigate other constructs that may affect sustained engagement and extend this work. These results have a large impact on the design process for serious games. Specifically, we acknowledge the roles of the factors stated above on adherence and engagement, and thus serious game designers will need to apply mechanics that cater to these constructs and game user researchers will need to test for them.

## ACKNOWLEDGMENTS

We would like to thank Aetna foundation and Northeastern University for supporting this work. We would also like to thank Lisa Andres the CEO of Igniteplay for her tremendous support for this project. Without the funds and the company support this work would not have been possible.